\documentclass[useAMS,usenatbib]{mn2e}

\usepackage{graphicx}
\usepackage{gensymb}
\usepackage{epstopdf}
\usepackage{rotating}
\usepackage{lscape}
\usepackage{aas_macros}
\usepackage{hyperref}
\usepackage{xspace}
\usepackage{dsfont}

\newcommand{\unit}[1]{\ensuremath{\, \mathrm{#1}}}

\def\apjl{ApJ}                % Astrophysical Journal, Letters
\def\apjs{ApJS}               % Astrophysical Journal, Supplement
\def\aap{A\&A}                % Astronomy and Astrophysics
\begin{document}

\title[]{Memory-efficient $w$-projection with the fast Gauss transform}
\author[K. W. Bannister et al.]{K. W. Bannister$^{1,2,3}$\thanks{E-mail:
keith.bannister@csiro.au}, T. J. Cornwell$^1$\\
$^{1}$CSIRO Astronomy and Space Science, PO Box 76, Epping NSW 1710, Australia  \\
$^{2}$Bolton Fellow \\
}

\date{Accepted 2011 XXX XXX. Received XXX December XX; in original form XXX XXX XXX}
\pagerange{\pageref{firstpage}--\pageref{lastpage}} \pubyear{2011}

\maketitle

\label{firstpage}

\begin{abstract}
We describe a method performing $w$-projection using the fast Gauss transform of \citet{Strain91}. We derive the theoretical performance, and simulate the actual performance for a range of $w$ for a canonical array. While our implementation is dominated by overheads, we argue that this approach could for the basis of a higher-performing algorithms with particular application to the Square Kilometre Array.
\end{abstract}


\section{Introduction}
Interferometers are composed of an array of antennas arranged in a 3 dimensional volume. For long observations, even instantaneously planar arrays have baselines that are non-coplanar due to Earth rotation synthesis. Correcting non-coplanar baselines in interferometers require the use of a de-projection technique to compensate for the non-coplanarity. To date, $w$-projection \citep{Cornwell08} is the most computationally efficient algorithm known.

Interferometric images are usually formed by convolutional resampling of the measured visibilities on a regularly sampled, two dimensional $uv$ plane. Convolutional resampling operates by multiplying each visibility by a convolution function and adding the result to the $uv$-plane. In the simplest case, the convolution function is used for anti-aliasing purposes, although it can be used to compensate for primary beam affects, or in the case of $w$-projection, to compensate for non-coplanar baselines. $w$-projection uses a convolution function that is dependant on $w$, which is the convolution of an anti-aliasing function, and a Fresnel term.

While more efficient that other methods \citep{Cornwell08}, $w$-projection does have some shortcomings. Firstly, the convolution functions take time to compute, and for a short observation, can dominate the time to compute an image. Secondly, the amount of memory to store the convolution functions can be large, and can exhaust the capabilities of a computing node. Finally, multiply-add operation is generally memory-bandwidth bound, so that the central processing unit (CPU) spends the majority of the time waiting for the data to arrive, and relatively little time doing the multiply-add operation.

The memory constraints of $w$-projection are of particular concern. The trend in computing has been for memory bandwidth to increase much more slowly than arithmetic capacity, which has roughly doubled every 18 months (Moore's law), with this trend likely to continue for the foreseeable future. Therefore, the performance of memory-bandwidth bound algorithms such as $w$-projection will not improve as quickly as an algorithm which is bound by arithmetic capacity.  Furthermore, modern high performance computers rely on having a very large number of small nodes, with each node having relatively limited memory. The requirement to store large convolution functions is, therefore, at odds with having nodes with small amounts of memory.

Memory efficient algorithms, therefore, will be an important step on the path to implementing wide-field imaging capabilities for the Square Kilometre Array (SKA). We have embarked on an program to research new approaches to interferometric imaging, with particular emphasis on algorithms that align more closely with the memory-bound computing that will be available in the time frame of the SKA.

In this paper we describe a different approach to $w$-projection. The key to our method is to represent the convolution function, including the $w$-projection term, and an anti-aliasing function, as a complex Gaussian. The gridding and degridding problems can then be solved using the variable-width fast Gauss transform (FGT; \citet{Strain91}). One advantage of this approach is that it does not require convolution functions to be computed and stored, and the theoretical memory bandwidth is less than standard gridding, in some situations.

In section \ref{sec:gauss_wproj} we describe Gaussian anti-aliasing functions used for $w$ projection and in section \ref{sec:wproj} we apply these functions with the Fast Gauss Transform, to the gridding and degridding problems. In section \ref{sec:performance} we derive the theoretical operations and memory requirements, and in section \ref{sec:perf_comparison} we compare the theoretical requirements for our method and the standard methods. In section \ref{sec:implementation} we compare the theoretical results with a real-world implementation. We discuss our results in \ref{sec:discussion} and we draw our conclusions in section \ref{sec:conclusions}.

\section{$w$-projection with a Gaussian anti-aliasing function}
\label{sec:gauss_wproj}

In this section we describe the gridding process, and use a Gaussian anti-aliasing function to obtain a closed form expression for the convolution function.

We acknowledge that the anti-aliasing properties of a Gaussian are not ideal. Typical interferometric imaging uses  prolate spheroidal wavefunctions, which have a optimal out-of-band rejection \citep{Schwab80}. Nonetheless, Gaussian functions have simple analytical relationships (e.g. the convolutions and Fourier transforms of Gaussians are both Gaussians) that allow us to proceed.

\subsection{Gridding}

The aim of convolutional gridding is to take visibilities sampled on arbitrary $u,v,w$ coordinates and interpolate them onto a regular grid so that a 2D fast Fourier transform (FFT) can be used to estimate the sky brightness distribution \citep[p. 134ff]{Taylor99}. Mathematically, the grid is evaluated at a point $(u_c, v_c)$ by multiplying each visibility by a shifted convolution function according to:

\begin{equation}
F(u_c, v_c) = \sum_{k=1}^{M}{C(u_c - u_k, v_c - v_k, w_k) V(u_k, v_k, w_k)} \label{eq:gridsum}
\end{equation}

\noindent where $M$ is the number of visibilities in to be gridded, $u$, $v$, and $w$ are the coordinates of the projected baseline (in wavelengths), $C(u, v, w)$ is a convolution function and  $V(u, v, w)$ is the measured visibility weight. Typically, $C$ is assumed to be zero outside some region of support (6--8 pixels). If we take no account for the $w$ term, then the convolution function is chosen to be an anti-aliasing function, and is independent of $w$, i.e. $C(u,v,w) = A(u,v)$. If we account for the $w$ term using $w$-projection of \citet{Cornwell08}, then the convolution function is given by

\begin{equation}
C(u, v, w) = A(u, v) * G(u, v, w) \label{eq:convfunc}
\end{equation}

\noindent where $*$ denotes convolution, $A(u, v)$ is an anti-aliasing function, and

\begin{equation}
G(u, v, w) = \frac{i}{w} \exp \left (-\pi i [u^2 + v^2]/w \right ) \label{eq:fresnel}
\end{equation}

\noindent is the Fresnel diffraction term required by the $w$ projection algorithm.

\subsection{Radially symmetric Gaussian anti-aliasing functions}

For a fixed $w$, equation \ref{eq:convfunc} is the convolution of the anti-aliasing function with a  complex Gaussian. If we choose a Gaussian for the anti-aliasing function $A$, the resulting convolution function is also Gaussian. To simplify notation, we will consider the radially symmetric problem by introducing a new parameter $t^2 = u^2 + v^2$, and can now write the Fresnel term as:

\begin{eqnarray}
G(t, w) & = & \frac{i}{w} \exp\left (- \frac{\pi i t^2}{w} \right ) \\
& = & \frac{i}{w} \exp \left ( \frac{-t^2}{ i \delta_G} \right )
\end{eqnarray}

\noindent with $\delta_G = - w/ \pi$. We can also write the Gaussian anti-aliasing function  as

\begin{equation}
A(t) = \exp{\left ( \frac{-t^2}{\delta_A}\right )}.
\end{equation}

The gridding function is given by the convolution $G(t) * A(t)$, which is another Gaussian, whose variance is equal to the sum of the variances of the original Gaussians, i.e., to within a scaling factor:

\begin{eqnarray}
C(t, w) & = & A(t) * G(t, w) \\
& = & \frac{i}{w} \exp \left ( - t^2\frac{1}{\delta_{A} + i \delta_{G}} \right ) \\
& = & \frac{i}{w} \exp \left ( - t^2 \frac{\delta_{A}}{\delta^2_{A} + \delta^2_{G}} \right )
\exp \left (  -t^2 \frac{ - i \delta_{G}}{\delta^2_{A} + \delta^2_{G}} \right ) \label{eq:prodconv}\\
& = &  \frac{i}{w} \underbrace{\exp \left ( - \frac{t^2}{\delta_R} \right ) }_\textrm{real envelope}
\underbrace{ \exp \left (  - \frac{i t^2}{ \delta_I} \right )}_\textrm{complex chirp} \label{eq:conv} \\
\end{eqnarray}

The width of the envelope is given by
\begin{equation}
\delta_R = \frac{\delta_A^2 + \delta_G^2}{\delta_A}, \label{eq:deltar}
\end{equation}

\noindent and width of the complex chirp is

\begin{equation}
\delta_I = \frac{\delta_A^2 + \delta_G^2}{\delta_G}.
\end{equation}

The convolution function is, therefore, a complex Gaussian, which can be expressed as the product of  real-valued Gaussian envelope (the $\delta_R$ term in equation \ref{eq:conv}) with a complex chirp (the $\delta_I$ term). Clearly, as $|w|$ increases, the width of the Gaussian envelope $\delta_R$ increases.

For a given visibility indexed by $j$ the form of the convolution function can be calculated from its $w$ coordinate using:
\begin{equation}
C(t, w_j) = \frac{i}{w_j} \exp{\left ( \frac{-t^2}{\delta_j} \right )} \label{eq:conv_ctw}
\end{equation}

\noindent where
\begin{equation}
\delta_j = \delta_A - i \frac{w_j}{\pi} \label{eq:conv_variance}
\end{equation}

\noindent is the width of the relevant convolution function, and $w_j$ is the $w$ coordinate of the visibility.

\subsection{Conversion from wavelengths to pixels}

Up to this point, the $u,v,w$ co-ordinates and the Gaussian widths have been in units of wavelength. For the remainder of this paper, width will often be quoted in terms of pixels. We use the terms pixel and $uv$-cell interchangeably. To convert from wavelengths to pixels, a size of $uv$-cell is required. The size of the $uv$-cell, (in wavelengths) is set by the inverse of the desired field of view (in radians) i.e. $\phi = 1/\theta_{\rm f.o.v}$.

Therefore, a Gaussian width in pixels can be calculated from the width in wavelength by:

\begin{equation}
\delta (\rm pixels) = \frac{\delta (\rm wavelengths)}{\phi^2}
\end{equation}

\section{$w$ projection with the Fast Gauss Transform}
\label{sec:wproj}

\subsection{A brief introduction to the Fast Gauss Transform}

In this section we introduce the Fast Gauss Transform (FGT) with variable scales \citep{Strain91} (hereafter S91), which is an approximate technique for calculating the sum of Gaussians. We conform to the language of S91, which uses the term `source' to describe a Gaussian function with a given position, amplitude and width. We will describe the application of the FGT onto the gridding problem in section \ref{sec:gridding}.

\begin{figure}
\includegraphics{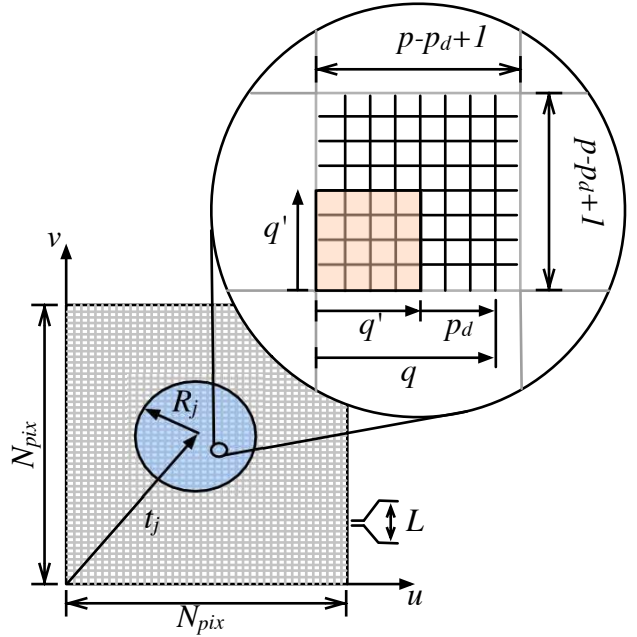}
\caption{This figure illustrates the process of gridding or degridding a single visibility. The $uv$ plane of size $N_{\rm pix}$ is partitioned into boxes size $L$. The box size can be the same size as a $uv$ cell. Each box contains $(p - p_d+1)^2$ Taylor coefficients. To grid or degrid a visibility indexed by $j$, all boxes within $R_j$ of the  visibility $uv$ position ($t_j$) are updated (circular blue shaded region). Within each box, $q^2$ Taylor coefficients are updated with no cheating. If cheating is enabled (i.e. $p_d > 0$), then only $q'^2$ of the $(p-p_d+1)^2$ Taylor coefficients (orange shaded region) are updated. $R_j$ has been exaggerated  for clarity.}
\label{fig:grid}
\end{figure}

The FGT works  by partitioning the evaluation region into square boxes, with each box containing a set of Taylor coefficients (Fig. \ref{fig:grid}). For each Gaussian `source', the Taylor coefficients for the boxes surrounding the source position are updated, with the size of the updated region being defined by the `width' of the source, i.e. wider Gaussians update more boxes. When all sources have been applied, the Taylor coefficients are evaluated at any number of target locations. The algorithm is parameterised by three numbers: $r$, which sets the box size,  $p$, which defines number of Taylor coefficients to consider, and $\delta$ which sets the smallest size of Gaussian that will be considered. Reducing the box size, or increasing the number of Taylor coefficients increases the accuracy of the sum.

\subsection{Gridding with the Fast Gauss Transform }
\label{sec:gridding}

Now that we have a Gaussian convolution function (Equation \ref{eq:conv_ctw}), we can solve the sum of Equation \ref{eq:gridsum} the FGT with \emph{source-dependant} scales. In the language of gridding visibilities, the evaluation region is the $uv$ plane. The `sources' are the visibilities, with the source position being given by the coordinates of the visibility on the $uv$ plane. The width of the source is related to the $w$ coordinate of the visibility as in Equation \ref{eq:conv_variance}, with the minimum width when $w=0$, being essentially the width of the anti-aliasing function. When all the visibilities have been applied to the Taylor coefficients, the Taylor coefficients are evaluated on a regular grid, for each $uv$ cell on the $uv$ plane.

For a detailed description of the algorithm, we refer the reader to S91. For the purposes of explaining our implementation, we outline the algorithm in the following  text, using similar notation to S91. This description applies to the one-dimensional case, but S91 describes the relationships to extend it to the multi-dimensional case (using multi-index notation), for which the 2D case is relevant here. Additionally, this description is only valid for real Gaussians. The extension to complex Gaussians will be considered in the next section.

The algorithm proceeds as follows:
\begin{enumerate}

\item Given values of $r$ and $p$ calculate:

\begin{equation}
\epsilon  =  \frac{r^p_p}{1 - r_p}, \label{eq:epsilon}
\end{equation}

\noindent where
\begin{eqnarray}
r_p & = & r\sqrt{\frac{e}{p+1}} < 1
\end{eqnarray}

\item Partition the $uv$ plane into the set of boxes $B$ with centers $t_B$, and side length $L = r\sqrt{2 \delta}$ parallel to the axes, where $\delta = \min_j \delta_j = \delta_A$ is the smallest Gaussian width to be considered.

\item For each visibility indexed by $j$, calculate the range $R_j = \sqrt{- \delta_j \log \epsilon}$. Also calculate the order $q \le p$ satisfying

\begin{equation}
\frac{r^q_{q,j}}{1 - r_{q,j}} \le \epsilon \label{eq:qerror}
\end{equation}

\noindent where $r_{q,j} = r\sqrt{e \delta/\delta_j (q + 1)}$.

\item For each box within $R_j$ of the visibility position $t_j$, update the Taylor coefficients according to:

\begin{equation}
C_\beta = \frac{1}{\beta !} \sum_{\delta_j \ge b} v_j \left (\frac{\delta}{\delta_j} \right )^{|\beta|/2} h_\beta \left( \frac{t_j - t_B}{\sqrt{\delta}} \right) \label{eq:cbeta}
\end{equation}

\noindent where $\beta$ runs from $0$ to $q$, $v_j$ is the visibility weight, and $h_\beta$ is the Hermite function (see S91).

\item Once all the visibilities have been applied, evaluate the Taylor coefficients at each cell in the $uv$ plane, by finding the box in which the $uv$ cell is situated, and calculating the sum:

\begin{equation}
F(t) = \sum_{\beta \le p} C_\beta \left( \frac{t_C - t_B} {\sqrt{\delta}}\right) ^\beta
\end{equation}

\noindent where $t_C$ is the position of the $uv$ cell, and $\beta$ runs from $0$ to $p$.

\end{enumerate}

The error in the evaluation of the sum is bounded by:
\begin{equation}
E_p   \le  V \frac{r^p_p}{1 - r_p},
\end{equation}

\noindent where
\begin{eqnarray}
V = \sum_{j=1}^{M} |v_j|.
\end{eqnarray}

$r$ and $p$ are chosen to make $|E_p| \le V \epsilon$.

\subsection{Extension to complex Gaussians}

The formulation of S91 is explicitly for real Gaussians, i.e. with $\delta_j, v_j \in \Re$, but in our case $\delta_j, v_j \in \mathds{C}$. A full derivation of the algorithm for complex values is outside the scope of this work . The important questions are whether the series of Equation \ref{eq:cbeta} converges (and for what values of a complex $\delta$) and what is a suitable range ($R_j$).

Empirically, we have determined that:

\begin{enumerate}

\item Using a range of $R_j = \sqrt{-\delta_{R,j} \log \epsilon}$ provides good results. This also makes sense because it is the amplitude of the envelope that determines the range.

\item For calculating the order $q$ (Equation \ref{eq:qerror}) we used
\begin{equation}
\delta_j = max(\delta_A, \delta_I) \label{eq:delta_error_complex}
\end{equation}

Other possibilities (e.g. $\delta_j = \delta_R$) reduced $q$ too aggressively as $w$ increased, which resulted in very large errors.

\end{enumerate}

In our case, the minimum value of $\delta$ occurs when $w=0$ which also coincides with $\Im(\delta) = 0$ so the box size is easily set as equal to $L = r \sqrt{2 \delta_A}$.

\subsection{Optimisations}
\label{sec:optimisations}

S91 describes a number of optimisations that can be used in the FGT with source-dependent scales:

\begin{enumerate}
\item S91 introduces a single set of Taylor coefficients, positioned at the center of the $uv$ plane, to capture Gaussians with the largest scales. Visibilities larger than an upper scale size $\delta_j > b$ can be added to this single set, rather than the sets associated with the boxes. S91 recommends the upper scale size be at least 10\% of the size of the evaluation region, in order to allow for practical precision in evaluating the large number of Taylor coefficients. In practice, most array configurations would have relatively few baselines with such a large $w$ that their convolution functions would span $>$ 10\% of the $uv$ plane, so this optimisation has limited impact.

\item S91 introduces a lower scale $a$ below which the Gaussian sum is directly evaluated. If there are only a few Gaussians with scales smaller than $a$, we can increase the box size, which reduces the memory and arithmetic capacity required, while incurring minimal penalty from direct evaluation. In practice, any interferometric observation is likely to be arranged such that the mean value of $w$ is approximately zero. Therefore, there are unlikely to be outlying, small Gaussians that would allow for increasing the box size. Once again, this optimisation has limited impact.

\item S91 proposes doing direct evaluation for boxes containing only a few evaluation points. As we are evaluating on a regular grid, we have the same number of evaluation points inside each box, so applying this optimisation would degenerate into standard gridding.

\item S91 proposes a method for avoiding needlessly accurate computation, i.e. for large scales (which are smoother than small ones) more boxes are updated, but smaller number of Taylor coefficients (i.e. the order $q$ described in Section \ref{sec:gridding}) per box can be updated. This optimisation \emph{is} useful and derives the most benefit for visibilities with large scales ($|w|$ large, see section \ref{sec:performance}.). We have incorporated this optimisation in Equation \ref{eq:qerror}.

\item S91 notes that the equation to calculate the order (Equation \ref{eq:qerror}) overestimates the error by at sometimes two orders of magnitude. Memory bandwidth and operations can be saved by `cheating' on the value of $q$ by decrementing $q$ by some value. This strategy is described in the section \ref{sec:cheating}.

\item The fact that each uv-cell contains a number of Taylor terms affords an additional axis available for parallelisation. The strategy is described in section \ref{sec:parallelisation}.

\end{enumerate}

\subsubsection{Cheating}
\label{sec:cheating}
There are two sources of error when a sample is gridded (see Figure~\ref{fig:error_pattern}). The first is from truncating the support; that is, only boxes only within a finite range $R_j = \sqrt{- \delta_R \log \epsilon}$ are updated. Convolutional gridding suffers from the same type of truncation error. The second source of  error is from truncation of the Taylor terms; that is, only $q < \infty$ Taylor terms are updated. S91 notes that the equation for calculating the error from truncation of  Taylor terms (Equation \ref{eq:qerror}) overestimates the error by several orders of magnitude. If this is the case,  for a given value of $\epsilon$, the error will be dominated by the truncated support, and the contribution from the truncation of the Taylor terms will be negligible. In principle, therefore, one can reduce the number of Taylor terms without substantially increasing the overall error, as long as it remains less than the error from the truncation of support.

In order to balance the errors from both effects, we introduce an additional parameter to the algorithm $p_d$, which decrements value of $q$, so that the actual value of $q$ that is used is $q' = \rm{max}(q - p_d, 0)$. This reduces the CPU and memory requirements of updating a given box, at the expense of increasing the errors due to truncating the Taylor terms.

\subsubsection{Parallelisation}
\label{sec:parallelisation}

One interesting property of our approach is that affords an additional axis of parallelisation (the Taylor terms) in addition to those available for standard gridding (typically, data parallelisation over frequency channels). We assume that the parallelism is achieved through message passing.

In section \ref{sec:performance} we show that the memory bandwidth and operations count for the FGT is dominated by updating the Taylor terms. As each of the Taylor terms is independent, one can in principle store each of the $(p-p_d + 1)^2$ Taylor terms on (up to) as many nodes, thereby dividing the per-node storage requirement by the number of nodes. For the case where $q'^2$ is a multiple of the number of nodes, the work is spread equally among the nodes. Each node can update its Taylor terms for the boxes within range, and, the total memory bandwidth is increased by the number of nodes.

One penalty of parallelising in this way is that each node must calculate the same interim values (e.g. the values of the Hermite polynomials). This duplication can be reduced by partitioning the Taylor terms among fewer nodes, with each node being responsible for a particular row or column of the Taylor terms.

If $q'^2$ is less than the number of nodes, then some nodes will have no work to perform, as they are responsible for Taylor coefficients that are not being updated. One must be careful, therefore, to choose an algorithm parameterisation i.e. ($r$,$p$ and $p_d$) that guarantees never to give $q'^2$ less than the number of nodes.

The efficiency of this approach depends crucially on the properties of the computing nodes. It is useful if the nodes are storage or memory bound, but not if the cost of computing the interim values dominates the computing time. As this tradeoff requires intimate knowledge of the particular computing architecture, we will not pursue a detailed analysis of parallelisation in this paper.

\subsection{Degridding with the Fast Gauss Transform}

Degridding is used as part of the major cycle of Cotton-schwab clean \citep{Rau11}. Degridding involves taking the dot-product of the $uv$ plane with the convolution function shifted to the location of a visibility at an arbitrary location (i.e. not at the center of a $uv$ cell). It is the dual of gridding.

Degridding can be performed analogously to gridding using the fast Gauss transform with \emph{target}-dependent scales (S91). All the same comments about optimisations for gridding (Section \ref{sec:optimisations}) apply as for degridding.

\subsection{Key differences between classical gridding and the FGT}

The key conceptual differences between classical gridding and FGT gridding are described in Table \ref{tab:vs}
\begin{table*}
\centering
\renewcommand{\arraystretch}{2}
\caption{Conceptual differences between classical gridding and gridding with the Fast Gauss Transform.}
\begin{tabular}{@{} l p{5cm} p{6cm} @{}} % Column formatting, @{} suppresses leading/trailing space
\hline
Property & Classical Gridding & Fast Gauss Transform \\
\hline
Accuracy: & Exact & Approximate, but with controllable error. \\
Anti-aliasing function: & Arbitrary. & Gaussian only. \\
Fundamental scale: & A $uv$-cell, set by imaging geometry. & A box, which can be smaller, or larger than a $uv$-cell. \\
Gridding a single visibility updates: & Cells on the $uv$-plane. & Taylor coefficients in a set of boxes. The cells of the $uv$-plane is calculated as a post-processing step. \\

\hline
\end{tabular}
\label{tab:vs}
\end{table*}

\section{Theoretical performance}
\label{sec:performance}

Here we derive the theoretical performance of the gridding and degridding processes using the fast Gauss transform, and traditional convolutional gridding.

\subsection{Operations count of the fast Gauss transform}

\subsubsection{Gridding}

To grid a single visibility of width $\delta_j$, the number of boxes that are within a circular region within the range $R_j = \sqrt{-\delta_R \log \epsilon}$ is approximately:
\begin{equation}
N_B = \frac{\pi}{4}(2 \lceil R_j/L \rceil + 1)^2 .
\end{equation}

The the order $q \le p$ is chosen to satisfy the error estimate:

\begin{equation}
\frac{r^q_{qj}}{1 - r_{qj}} \le \epsilon
\end{equation}

\noindent where $r_qj = r\sqrt{e \delta/\delta_j(q + 1)}$.

To apply cheating (Section \ref{sec:cheating}), we decrement the value of $q$, such that $q' = \rm{max}(q - p_d, 0)$.

Now we consider work required to update the Taylor coefficients in each box (Eq. \ref{eq:cbeta}). The key is to compute and store the Hermite functions, and powers of $(\delta/\delta_j)^{|\beta|/2}$ separately, and perform the sum at the end, by taking advantage of the product form of the multi-index notation for multiple dimensions.

The Hermite function is a polynomial multiplied by a complex exponential. The argument for the complex exponential is the same for every order $q'$, so only one calculation is required per set of Hermite function evaluations. The way in which  a complex exponential is computed by a CPU is highly implementation-dependant, so we simplify the analysis here and assume that it requires $N_{\rm cexp}$ floating point operations.

We observe that the Hermite function of order $q'$ has only $N_c=\lfloor q'/2 \rfloor + 1$ nonzero coefficients. Second, we note that we will evaluate the Hermite functions for order $0 \le \beta \le q'$ with the same argument. Therefore, we can calculate the powers of the argument initially with $q'$ operations. Evaluating each Hermite function requires only  $N_c$ additional operations. Therefore, to compute all Hermite functions up to order $q'$ requires approximately:
\begin{eqnarray}
N_{\rm ops, hermite} & = & N_{\rm cexp} + q' + 2 \sum_{\beta = 0}^{N_c} N_c \nonumber \\
& = & N_{\rm cexp} + q' + 2 \sum_{\beta = 0}^{N_c} \lfloor \beta/2 \rfloor + 1 \nonumber \\
& \simeq & N_{\rm cexp} + q' + N_C (N_C + 1)
\end{eqnarray}

The multi-index powers of $t_\beta = (\delta/\delta_j)^{|\beta|/2}$ can be computed recursively, by computing  $t_n = (\delta/\delta_j) t_{n-2}$ with $t_0=(\delta/\delta_j)^{1/2}$, and $t_1=(\delta/\delta_j)^1$. This requires $2(q'+1)$ operations $0 \le n \le 2q'$.

The weight $v_j$ can be folded into the power values also, with $2(q'+1)$ operations.

Finally, the full sum to update $C_\beta$ over two dimensions requires the multiplication of Hermite functions for 2 dimensions, and the powers, plus the accumulation, which requires $3 (q'+1)^2$ operations in total.

The total number of operations to update a single box is therefore the sum of two Hermite evaluations (one for each dimension), plus the powers of $\delta/\delta_j$, the weights and the final sum, i.e.:

\begin{equation}
N_{\rm grid} = 2 (N_{\rm cexp} + q' + N_C[N_C + 1]) + 4(q'+1) + 3(q'+1)^2
\end{equation}

For small $q'$, the cost is dominated by the cost of evaluating the complex exponential. For large $q'$, it is dominated by the multiply-add step in the final sum.

\subsubsection{Degridding}

The operations count for the degridding process is similar to the gridding case. The order $q'$ is evaluated identically, as are the Hermite functions and the powers of $\delta/\delta_j$. The only difference is that no weight is included, and the multiply-add stage includes the product of the two hermite functions, the power and the Taylor term, requiring $4(q'+1)^2$ operations.

The number of operations per visibility is therefore:
\begin{equation}
N_{\rm degrid} = 2(N_{\rm cexp} + q' + N_C[N_C + 1]) + 2(q'+1) + 4(q'+1)^2.
\end{equation}

Once again, for small $q'$, the cost is dominated by the cost of evaluating the complex exponential. For large $q'$, it is dominated by the multiply-add step in the final sum.

\subsubsection{Memory bandwidth}

Calculating memory bandwidth is complicated by the issue of cache hierarchy. We assume the simplest case: i.e. no caching. In practice, this is a reasonable assumption, as visibilities are often stored in no particular order. Therefore, the desired grid (and convolution functions in the case of convolutional gridding) are essentially random, meaning that all memory accesses go to the main memory and bypass the caches.

For the fast Gauss transform, gridding a visibility requires the updating of $(q'+1)^2 N_B$ complex coefficients. The coefficients must be read into the processor, updated, and written back to memory, requiring $2 (q'+1)^2 N_B$ memory transactions per visibility.

For degridding, the coefficients do not need to be written back to memory, therefore the only $(q'+1)^2 N_B$ memory transactions are required.

\subsubsection{Storage}

For an image size of $N_{\rm pix}^2$, the gridding operation requires storage for $(N_{\rm pix}/L)^2 (p- p_d+1)^2$ complex coefficients. Degridding requires the same storage.

\subsection{Performance of convolutional gridding and degridding}

In this section we derive the operations count, memory bandwidth and storage required to perform gridding and degridding with stored convolution funcitons. In order to put the two methods on equal footing in terms of imaging performance, we will use Gaussian anti-aliasing functions for the $w$-projection, rather than the commonly-used prolate spheroidal wavefunctions. This means that the imaging performance of the two approaches is essentially the same (except for the errors in truncating the Taylor series of the FGT) and the support size of the convolution functions is also easily computed.

We assume standard $w$ projection computes the convolution functions in advance and stores them in memory. We assume a convolution function of 1-D size for the $k$th $w$-plane $M_{\rm k} = k \Delta w$, where $k$ is an integer, and $\Delta w$ difference in size between different $w$ planes. Usually the cached convolution function is oversampled by a factor $\kappa$ of 4--8 in order to accurately grid visibilities whose coordinates are not exactly in the center of a $uv$ cell.

We assume that we truncate the convolution function when the real envelope reaches a value $\epsilon$, which is at a distance $t_\epsilon = \sqrt{- \delta_R \log \epsilon}$ from the centre of the convolution function. Therefore, the 1D support size in pixels, of the $k$th $w$-plane is (by substituting Equation \ref{eq:deltar})

\begin{eqnarray}
M_{k} & = &2 t_\epsilon \nonumber \\
& = & 2 \sqrt{\log \epsilon \left (\delta_A + \frac{k^2 \Delta w^2}{\pi^2 \delta_A}  \right )}
\end{eqnarray}

\subsubsection{Operations count}

Gridding a visibility requires two operations per point (weight times convolution function, add to grid), for a total of $2 M_{\rm k}^2$ operations per visibility. Degridding also requires two operations per point(convolution function times grid, add to result) and and therefore requires $2 M_{\rm k}^2$  operations per visibility.

\subsubsection{Memory bandwidth}

Gridding requires the convolution function, and the grid position to be retrieved and written back to memory, requiring $3 M_{k}^2$ memory transactions per visibility. Degridding has no requirement to write back the result, (as the intermediate sum is stored in a register), so only $2 M_{\rm k}^2$ memory transactions are required.

\subsubsection{Storage}

To compute the total memory required, we will determine the amount required to store $N_w$ $w$-planes for values of $w$ between $[0, +w_{\rm max}]$, uniformly sampled with width $\Delta w = w_{\rm max}/N_w$. In practice, we need to store $w$-planes for $[-w_{\rm max},0)$, so we will take the above result and multiply by two.

The amount of memory required to store the convolution functions is therefore equal to the sum of the oversampled convolution functions, i.e.:

\begin{eqnarray}
\frac{N_{\rm mem}}{2} & = & \sum_{k = 0}^{N_w}{(\kappa M_k)^2} \nonumber \\
& = & -4 \kappa^2 \log \epsilon \left (N_w \delta_A + \frac{\Delta w^2}{\pi^2 \delta_A} \frac{N_w}{6} [N_w + 1][2 N_w + 1] \right ) \\
& \simeq & -\frac{4}{3} \kappa^2 \log \epsilon \frac{w_{\rm max}^2}{\pi^2 \delta_A} N_w
\end{eqnarray}

\section{Theoretical Performance Comparison for Gridding}
\label{sec:perf_comparison}

We consider the gridding problem for two scenarios, $L=1$ and $L=2$. We  assuming $\delta_A=1$~pixel, a required accuracy of $\epsilon=10^{-3}$, a 1 degree field of view and a 6~km maximum baseline at $\lambda = 0.2$ giving a $w_{\rm max}$ of $30 k \lambda$. The convolution function size at maximum $w$ of $M_{\rm wmax}=191$ pixels. We assume the cost of evaluating a complex exponential is $N_{\rm cexp} = 100$ flops. This estimate is also justified by measurements of our implementation (Fig. \ref{fig:procratio_vs_w}).

\begin{figure}
\centering
\includegraphics[width=\linewidth]{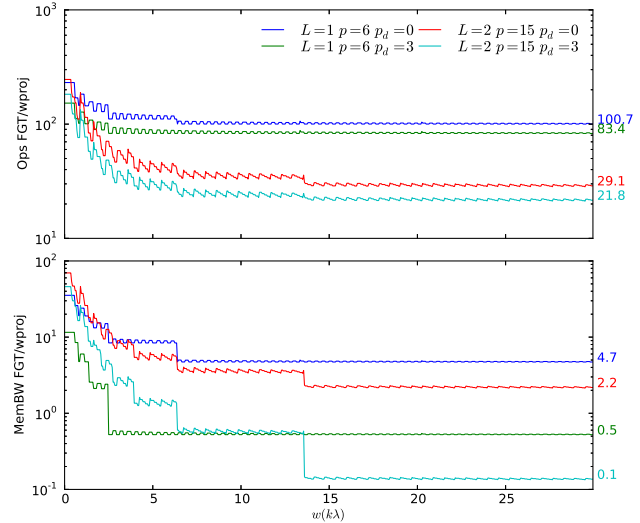}
\caption{Gridding with the fast Gauss transform can save memory bandwidth for large support sizes, at a cost of floating point operations. Plotted here is the ratio of operations (top panel) and memory bandwidth (bottom panel) for the fast Gauss transform compared with convolutional gridding (see Section \ref{sec:performance}). The x-axis is the value of the $w$ coordinate, which is also a measure of the width of the convolution function (Equation \ref{eq:conv_variance}). The target error for each algorithm was set to $\epsilon = 10^{-3}$. The fast Gauss transform was configured with $\delta_A=1$~pixel and we have assumed $N_{\rm cexp}=100$. The step drops in the ratio occur when the FGT reduces the order $q'$ for the Taylor truncation. For $w>14 \unit{k\lambda}$, the FGT with $L=2$ requires 21.8 times more operations but only 0.1 times the memory bandwidth of classical gridding, but only if cheating is enabled with $p_d=3$.}
\label{fig:fgt_vs_gridding}
\end{figure}

The FGT method for $L=1$ and $L=2$ is outperformed by standard gridding both in terms of operations and memory bandwidth, if cheating is not enabled (Figure \ref{fig:fgt_vs_gridding}). For the $L=1$ case in particular, the operations count is dominated by the complex exponential.

If cheating is enabled, then the situation is substantially improved. Once the $w$ value reaches above a certain threshold, the order $q' = 0$, and only one Taylor coefficient is updated per box. In the $L=1$ case, the memory bandwidth is reduced to 0.5 of the standard gridding case, because the FGT only requires only two memory transactions per  pixel (read + write coefficient), while standard gridding requires three (read pixel + read convolution function + write pixel). For $L=2$ the improvement in memory bandwidth is even more with the FGT requiring 10 times less memory bandwidth than standard gridding. The FGT is more efficient as it only updates one coefficient for each box, which encapsulates 4 pixels. The FGT case is also improved because it only updates boxes within a circular region, while standard gridding updates all pixels within a square.

This encouraging result leads to a number of questions: is the FGT bound by memory bandwidth? Can these performance gains be realised in practice? Can cheating with $p_d=3$ give reasonable image errors?  We will address these questions in the following section.

\section{Implementation}
\label{sec:implementation}

We implemented the gridding and degridding algorithms as described in Section \ref{sec:wproj} in C++. As with the theoretical simulation, we set the width of the anti-aliasing function equal to $\delta_A=1$~$uv$ cell. We aim explore the parameter space similar to the theoretical analysis, i.e. around $L=1$ and $L=2$ and ranges of errors around $10^{-3} < \epsilon < 10^{-2}$.

To compare the gridding errors, we compared the relative error of a visibility gridded with the FGT with the equivalent complex Gaussian (Equation \ref{eq:prodconv}), calculated over $256^2$ pixels (which is larger than the 191 pixels we expect for $\epsilon = 10^{-3}$ and $w=w_{\rm max}$) . To compare the compute times, we compare the time to grid 100 visibilities of fixed $u,v,w$ coordinates with the FGT, with the same number of visibilities gridded with standard gridding. The support size of the gridding kernel was chosen to have with equivalent error to the FGT error, i.e. $M_k^2$ pixels. We chose 101 $w$ planes from $0$ to $w_{\rm max}$. The processing  was performed on an Intel Core 2 Duo processor running at 2.66~GHz, with 32~kB of L1 cache, 3~MB of L2 cache and with 8~GB of 1057~MHz DDR3 RAM.

We consider here only the results for the gridding operation, as the degridding algorithm results essentially the same.

\subsection{Error patterns}

\begin{figure*}
\centering
\includegraphics[height=14cm]{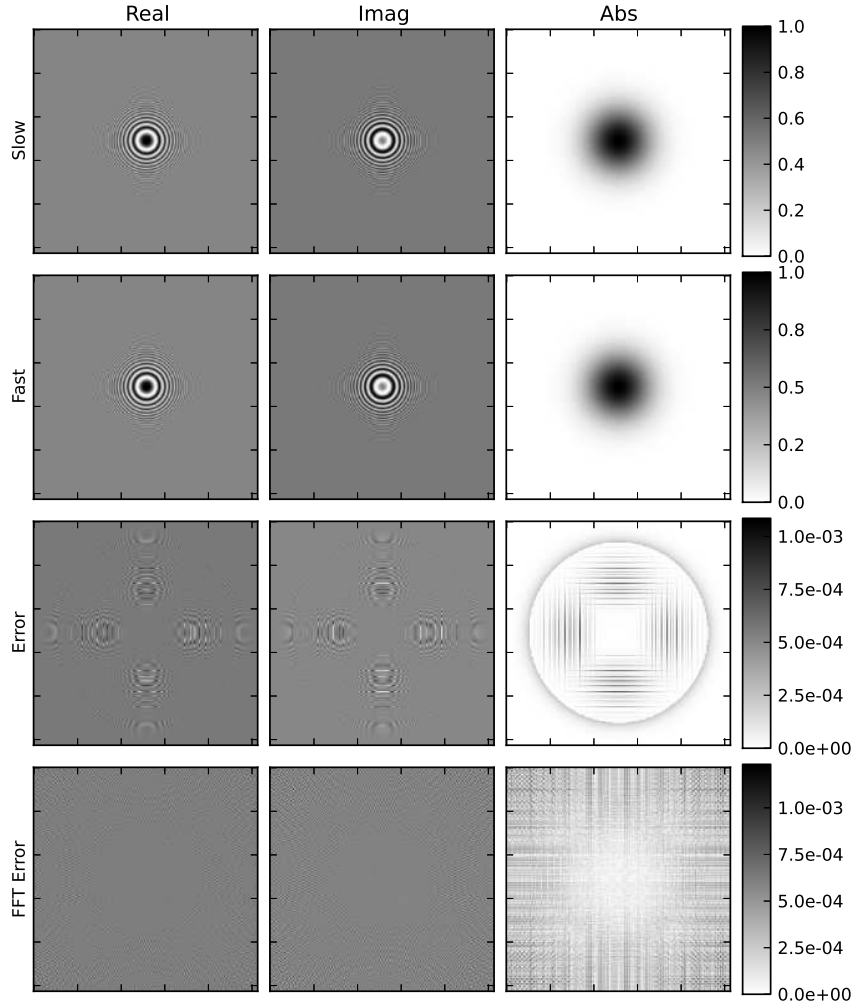}
\caption{Error pattern for a visibility gridded at $w=w_{\rm max}$ for $L=0.7$ and $q'=7$. Each box is 256x256 $uv$ pixels. The rows are the visibility computed with the full complex Gaussian, the FGT, the error, and the FFT of the error. The columns are the real, imaginary and absolute values of the data respectively. The circular component of the error is due to the truncated range, while the horizontal and vertical components of the error are due to the truncation of the Taylor series. This example was chosen because it illustrates both components clearly.}
\label{fig:error_pattern}
\end{figure*}

A typical error pattern is shown in Figure \ref{fig:error_pattern}. The absolute error in the gridded visibility contains two components. The circular component is due to the truncation to finite support, i.e. boxes outside the circle were not updated. The vertical and horizontal lines are due to truncation to a finite number of Taylor terms, with the largest error where the grid is evaluated near the boundary of adjacent boxes. The box structure of the gridding process can also be clearly seen in the FFT error, which also shows a box structure.

\subsection{Peak Image Errors vs $w$}

\begin{figure*}
\centering
\includegraphics[width=\linewidth]{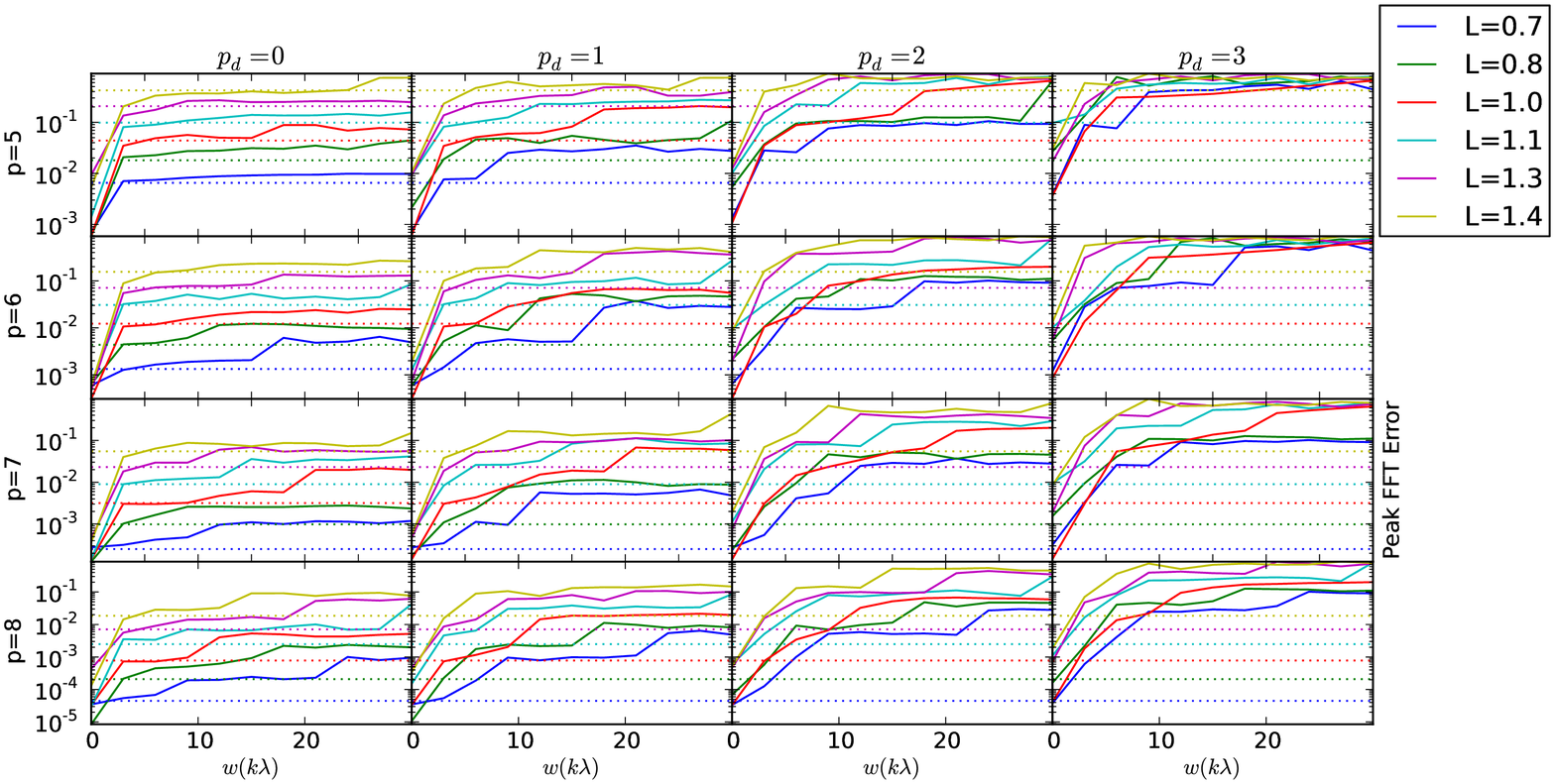}
\includegraphics[width=\linewidth]{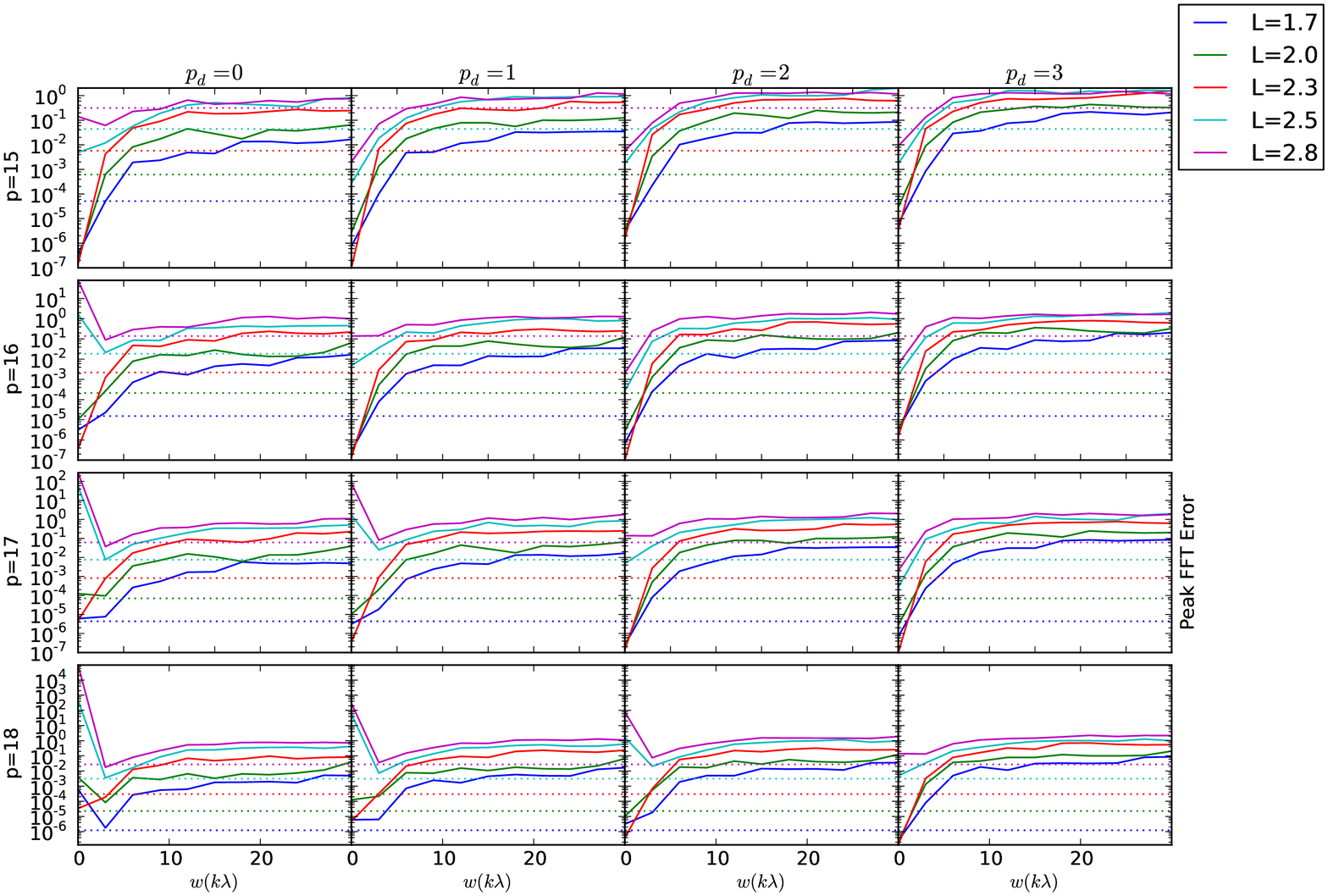}
\caption{The image plane error of the FGT is a strong function of $w$ and the algorithm parameterisations. Here we plot the relative error in the image plane of the FGT vs. $w$ for a range of algorithm parameterisations. The solid lines are the measured error in the image plane and the dotted lines are the predicted values of $\epsilon$  (Equation \ref{eq:epsilon}). The top panels are for a box size $L\sim1$, and the bottom panels are for a box size of $L\sim2$ pixels.}
\label{fig:error_vs_w}
\end{figure*}

The peak error in the FFT of the visibility (i.e. the image plane), as a function of the $w$ for  a range of algorithm parameterisations are shown in Figure~\ref{fig:error_vs_w}.  To begin with, we consider the case for $L~\sim 1$. Somewhat surprisingly, Equation~\ref{eq:epsilon}, which describes the error in gridding a single visibility on $uv$ plane, also predicts the error the image plane for many parameterisations and values of $w$. For $p_d=0$, and small $w$, the errors are substantially smaller than the predicted $\epsilon$. As $w$ increases, $q'$ is reduced, and the error generally matches the predicted $\epsilon$. For some parameterisations (e.g. $r=0.5$, $p=7$) there is a jump in error above predicted $\epsilon$ at a particular $w$.

For the $L~2$ case, for small $w$, the errors are at or below the theoretical limits in most cases, however in many cases the errors are considerably worse than for $L=1$. For example, for $w$ small in some cases (e.g. $L=2.5$, $p=18$, $p_d=0$) the error \emph{increases} far beyond the theoretical limit. For large $w$, the errors are several orders of magnitude worse than predicted.

\subsection{Processing time}

\begin{figure*}
\centering
\includegraphics[width=\linewidth]{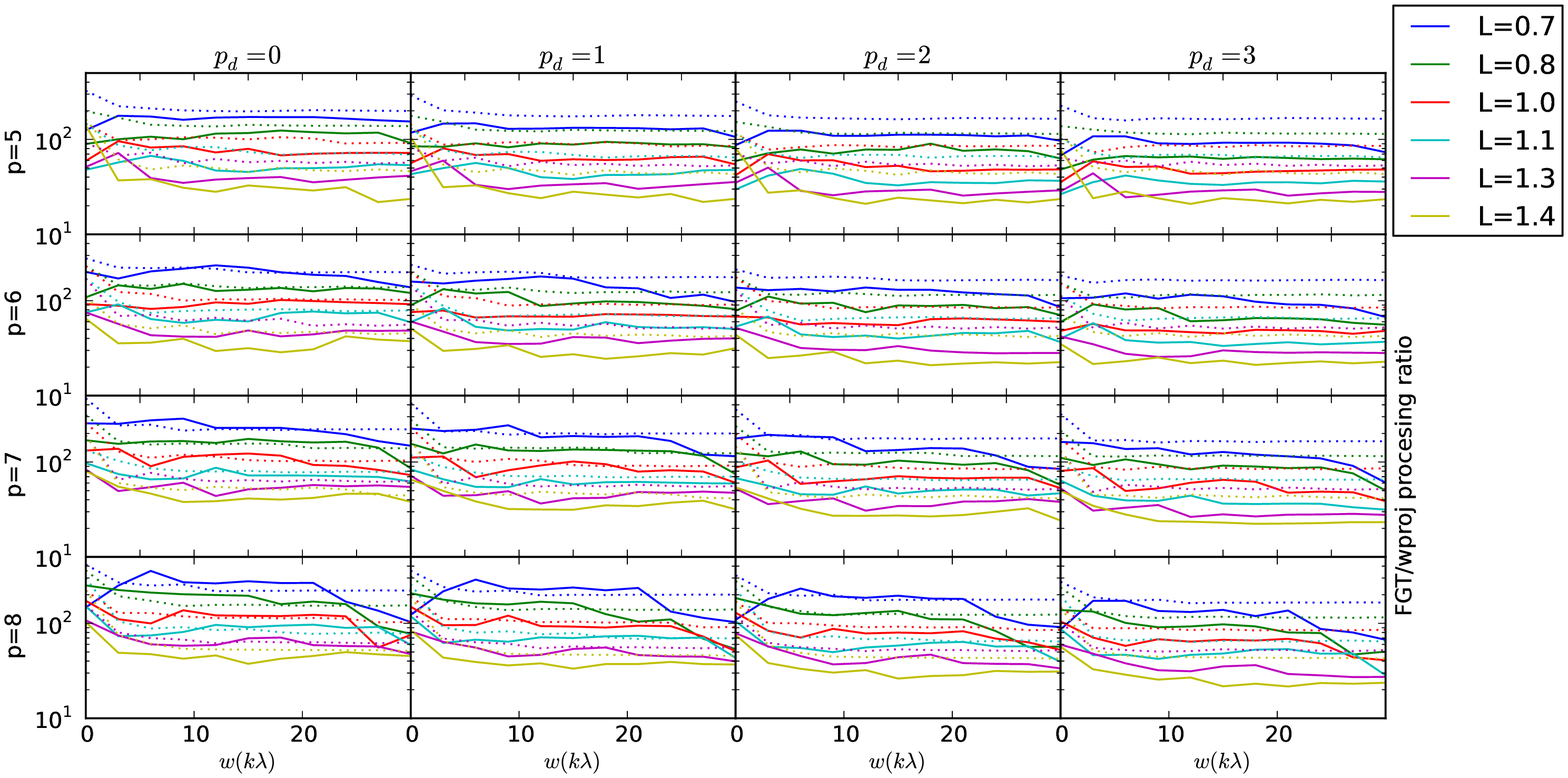}
\includegraphics[width=\linewidth]{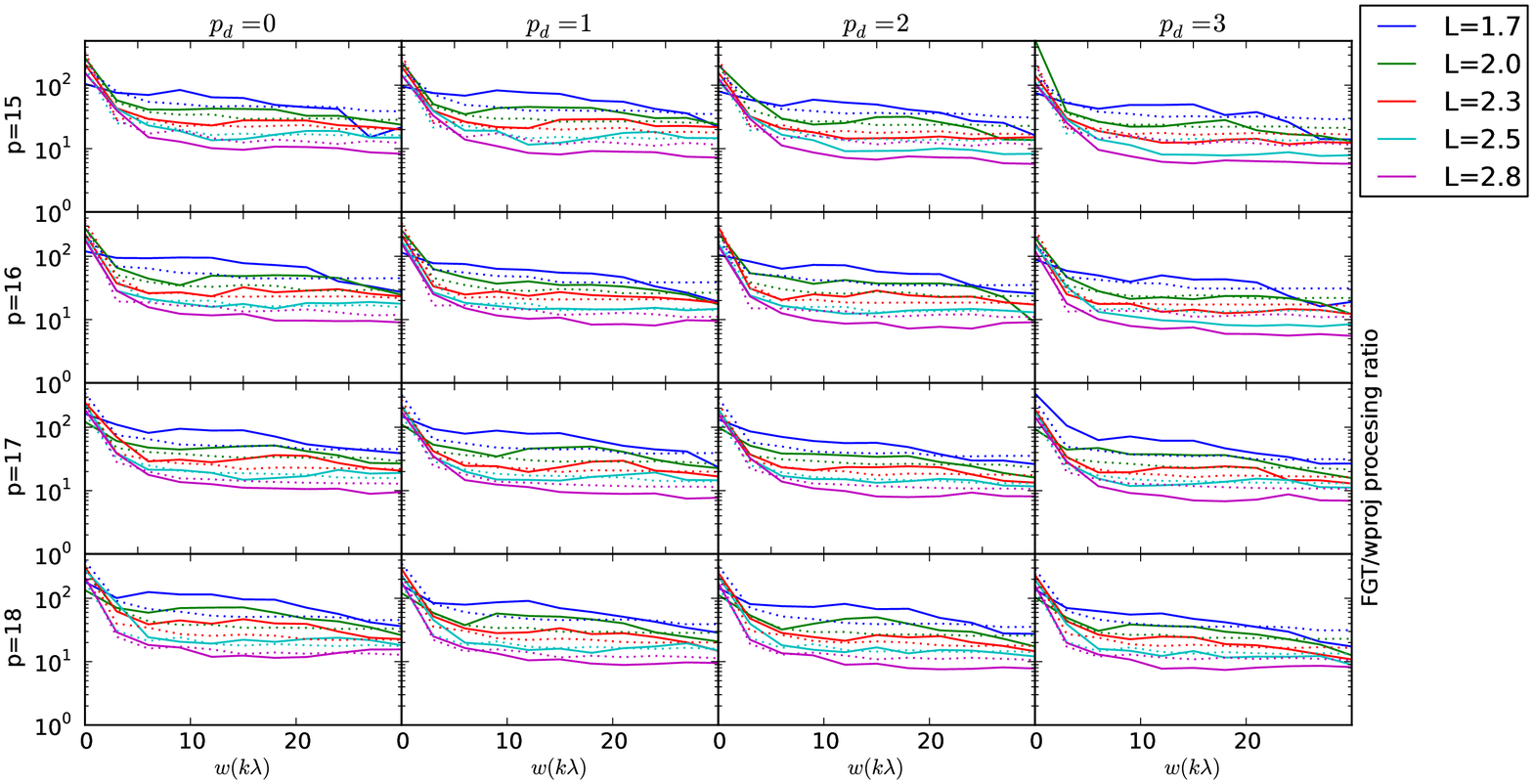}
\caption{Our implementation of gridding with  the fast Gauss transform is substantially slower than standard gridding, and compute bound. Here we plot the ratio of compute time to grid 100 visibilities with the FGT vs. standard gridding, over a range of $w$.  The solid lines are the ratio of the measured processing times, while the dotted lines are predicted ratio in operations count (the same as top bottom panel of Fig. \ref{fig:fgt_vs_gridding}). The top panels are for a box size $L\sim1$, and the bottom panels are for a box size of $L\sim2$ pixels.}
\label{fig:procratio_vs_w}
\end{figure*}

The measured processing time for the FGT vs. standard gridding is shown in Figure~\ref{fig:procratio_vs_w}. Our implementation operates substantially slower than standard gridding for all parameterisations and ranges of $w$, and the operations rate is well matched by the our theoretical model for the operations count.

\section{Discussion}
\label{sec:discussion}

\subsection{FGT Implementation}

The results from testing our implementation are somewhat conflicting. On the plus side, Figure \ref{fig:error_vs_w} clearly shows that, for small values of $w$, Equation~\ref{eq:qerror} is overestimating the truncation error, and the value of $q'$ being used is too large. This was the original motivation for introducing cheating, and it is clear that the error is still overestimated even for $p_d=3$ in some cases.

Unfortunately, for larger values of $w$, the fact that the error jumps above the predicted $\epsilon$, even  with no cheating, implies that our empirical extension to the FGT to complex Gaussians, (in particular Equation~\ref{eq:qerror}, and Equation \ref{eq:delta_error_complex}) is not correct. As a result of this incorrectness, our current cheating algorithm has limited value. Even with $p_d=1$, most parameterisations contain jumps in error of factors of few to 100 over the range of $w$. A different mapping between $w$ and $q'$ is required to maintain a constant error across $w$.

Increasing the box size above $L>2.3$ appears to have  catastrophic affects on the error for small $w$. We suspect this is due either to numerical instability in evaluating high-order polynomials, or more likely, the Hermite expansions simply fail to converge when the Gaussian width is less than some function of the box size.

Our implementation's processing time is clearly compute-bound  as our computing times match our theoretical model for the operations count almost exactly in most instances. The theoretical model suggests the operations count is dominated by the cost of computing the complex exponential, and we expect that is the case, although other overheads may also be responsible. Sadly, we were not able to approach the memory-bound performance that originally motivated this work.

\subsection{Future work}

In spite of the disappointing performance of our implementation, we hope that some of the ideas from this work could be extended  in future. In particular:

\begin{itemize}
\item The idea of having tuneable error is attractive. For arrays with high sidelobes in their synthesised beams (i.e. poor $uv$ coverage),  large gridding errors can be made, as long as they remain below the clean threshold for the given major cycle.

\item The required memory bandwidth of our algorithm is only weakly dependant on  $|w|$, so for arrays with very long baselines, our algorithm may be suitable.

\item The Hermite expansions were originally proposed by \citet{Greengard91} for real Gaussians. It may be that faster-converging expansions can be found for complex Gaussians - particularly in the case for large $w$ where the real envelope is smooth and the complex chirp contains many peaks.

\item The form of the convolution function is smooth at the centre and oscillates more rapidly as $t$ increases, with the fastest oscillations being damped at the edges by the envelope. Currently, our approach is to use the same order for all boxes, which leads to the largest errors being made away from the centre of the convolution function  (Figure~\ref{fig:error_pattern}) where the oscillations are the fastest and largest. There may, therefore, be some value in varying the order $q'$ as  a function of the distance from the centre, thereby applying the majority of the computational effort where the errors are likely to be worst.

\item We implemented the polynomial evaluations in a naive manner. More sophisticated methods exist  \citep[p. 485ff]{TAOCP2} that could reduce the operations count and improve numerical stability.

\item To calculate the complex exponential, we used the {\sc CEXP} function from the standard C library, which is accurate but slow. Faster, approximate methods are available\footnote{\url{http://gruntthepeon.free.fr/ssemath/} (e.g. \citep{Schroudolph98}}. For an implementation whose run time is dominated by {\sc CEXP} such as ours, these methods could improve speed by a factor of a few, albeit at the cost of some accuracy.

\item Gaussian anti-aliasing functions are not well suited to imaging, as the alias rejection is not very good. The modern state of the art is to use prolate spheroidal wavefunctions \citep{Schwab80}, that have better alias rejection. Unfortunately, the Hermite functions are not a suitable basis for Taylor expansion of the prolate spheroidal wavefunctions. If suitable analytic multipole expansions of the convolution of the complex chirp, and the prolate spheroidal wavefunctions can be found, then they can be applied with the fast multiple method \citep{Greengard87, Greengard88} to obtain a similar result as our FGT, but with better anti-aliasing properties.

\item The fast multipole method could also be used to compensate for the primary beam in the $uv$-plane (AW-projection, \citep{Bhatnagar08}). Once again, this would require analytic multipole expansions of the required convolution functions. While this may not be possible with an arbitrary primary beam, it may be more tractable if we assume Gaussian primary beams.

\item For problems where the compute time of $w$-projection is dominated by the time to compute convolution functions (by Fourier transforming a phase screen in the image plane), using standing gridding with Gaussian anti-aliasing functions could be preferable, as they can be directly and efficiently computed in the $uv$ domain.

\item Easily parallelised algorithms will clearly be important for large arrays such as SKA, where the computational work is likely to be spread over many thousands of nodes. Once the gridding problem has been distributed over the usual axes of frequency, pointing direction and polarisation, it is possible that  problem may still be too large to be efficiently computed by a single node. For the gridding operation, the option of distributing and parallelising over Taylor coefficients adds an additional axis which an be exploited when distributing work among processors, over and above the traditional axes.

\end{itemize}

\section{Conclusions}
\label{sec:conclusions}

We have described a procedure to perform $w$-projection with the fast Gauss transform with variable scales, where the anti-aliasing function is chosen to be a Gaussian. The gridding problem is solved by the FGT with source variable scales, and the degridding problem by the FGT with target variable scales. While the theoretical efficiency of our approach is encouraging, we were not able to approach the the theoretical performance gains with our implementation. Nonetheless, we find that $w$ projection with approximate algorithms such as the FGT or fast multipole methods may yet have promise, by having the attractive properties of tuneable error, an additional parallelisation axis, and no calculation and storage of convolution functions. The methods require additional research, to improve the practical implementation, find expansions with faster convergence, and find closed forms with better anti-aliasing properties.

\section{Acknowledgements}
The authors would like to thank Maxim Voronkov and Ben Humphries for invaluable help with C++ coding,  Matthew Whiting for the idea of tuneable errors within major cycles, and Oleg Smirnov for his insightful referee comments.

\bibliographystyle{scemnras}
\bibliography{Master.bib}

\label{lastpage}
\end{document}